\documentclass{PoS}

\title{Machine and deep learning techniques in heavy-ion collisions with ALICE}
\ShortTitle{Machine and deep learning techniques in heavy-ion collisions with ALICE} 

\author{\speaker{R\"udiger Haake} for the ALICE collaboration\\
        CERN\\
        E-mail: \email{ruediger.haake@cern.ch}}

\abstract{
Over the last years, machine learning tools have been successfully applied to a wealth of problems in high-energy physics.
A typical example is the classification of physics objects.
Supervised machine learning methods allow for significant improvements in classification problems by taking into account observable correlations and by learning the optimal selection from examples, e.g. from Monte Carlo simulations.
Even more promising is the usage of deep learning techniques. Methods like deep convolutional networks might be able to catch features from low-level parameters that are not exploited by default cut-based methods.

These ideas could be particularly beneficial for measurements in heavy-ion collisions, because of the very large multiplicities. Indeed, machine learning methods potentially
perform much better in systems with a large number of degrees of freedom compared to cut-based methods. Moreover, many key heavy-ion observables are most interesting at low transverse momentum where the underlying event is dominant and the signal-to-noise ratio is quite low.

In this work, recent developments of machine- and deep learning applications in heavy-ion collisions with ALICE will be presented, with focus on a deep learning-based b-jet tagging approach and the measurement of low-mass dielectrons. While the b-jet tagger is based on a mixture of shallow fully-connected and deep convolutional networks, the low-mass dielectron measurement uses gradient boosting and shallow neural networks. Both methods are very promising compared to default cut-based methods.
}

\FullConference{EPS-HEP 2017, European Physical Society conference on High Energy Physics\\
		5-12 July 2017\\
		Venice, Italy}

\begin{document}

\section{Introduction}
Machine learning techniques were already successfully applied to several analyses in high-energy physics. Depending on the analysis, they allow a better performance or can spare time required for finding optimal parameters.

The application of deep learning, which usually means the usage of many layers of neural networks, is relatively new to high-energy physics. There are, though, some examples also in jet physics\,\cite{JetImages}.

It is a promising approach, which typically uses basic, low-level parameters instead of engineered observables exploiting physics knowledge. It might in particular help with problems for which the correlations of low-level parameters are poorly understood.

This article will present two ALICE analyses that make use of machine and deep learning techniques: Deep-learning b-jet tagging and background suppression for low-mass dielectrons.

\section{b-jet tagging}

The Quark-Gluon Plasma (QGP) is the hot and dense medium that is formed in high-energy heavy-ion collisions. It is the main focus of study of the ALICE experiment.

Jets are powerful probes to understand the properties of the Quark-Gluon Plasma. In the final state, they can conceptually be described as collimated sprays of particles produced in a hard scattering in the initial state. Therefore, jets are an excellent tool to access a very early stage of the heavy-ion collision. The jet constituents represent the final state remnants of the fragmented partons that were scattered in the reaction. While all the detected particles have been created in a non-perturbative process (i.e. by hadronization), ideally, jets represent the kinematic properties of the originating partons. Thus, jets are mainly determined by perturbative processes due to the high-momentum transfer and their production cross sections can be calculated with pQCD. Even though this conceptual definition is descriptive and very simple, the technical analysis of those objects is complicated, in particular in the high-multiplicity environment of heavy-ion collisions.

Jets originating from heavy-flavor (beauty or charm) quarks are of particular interest. The parton energy loss of partons is different for heavy quarks: Compared to gluons, quarks in general are expected to lose less energy in the medium because of the smaller color charge. In addition, for massive quarks, gluon bremsstrahlung is suppressed at smaller angles with respect to the jet axis ("dead cone effect"\,\cite{DeadCone}). Therefore, heavy-flavor jets can help to understand the parton energy loss mechanism the medium and to assess the properties of the QGP.\\

The present analysis focuses on the identification and measurement of b-jets. In Monte Carlo simulations, which are used to train the classifier, the type of a jet is determined as follows.
\begin{itemize}
  \item A jet is tagged as a b-jet, if a hadron with beauty content is contained in the Monte Carlo history of the jet cone within a radius $R=0.4$.
  \item If instead a hadron with charm content is found, the jet is defined as a c-jet.
  \item Other jets are defined as light-flavor (u, d, s, gluon) jets.
\end{itemize}

B-hadrons decay in the (sub-)millimeter range around the primary vertex ($c\tau \approx 500 \mu\mathrm{m}$). This displacement is measurable with high-precision vertex detectors and serves as main physics property to discriminate b-jets. The displacement can be measured via secondary vertex reconstruction and track impact parameters. For each track, two impact parameters are defined as the track distance of closest approach to the primary vertex, either in the transverse or the longitudinal plan. The conventional approach is to cut on such parameters to extract a b-jet enriched sample.
Instead, the present approach applies deep-learning techniques to several low-level parameters.

\subsection{Model design and input features}

%%%%%%%%%%%%%%%%%%%%%%%%%%%%%%%%%
\begin{figure}[htp]
\includegraphics[width=0.7\textwidth]{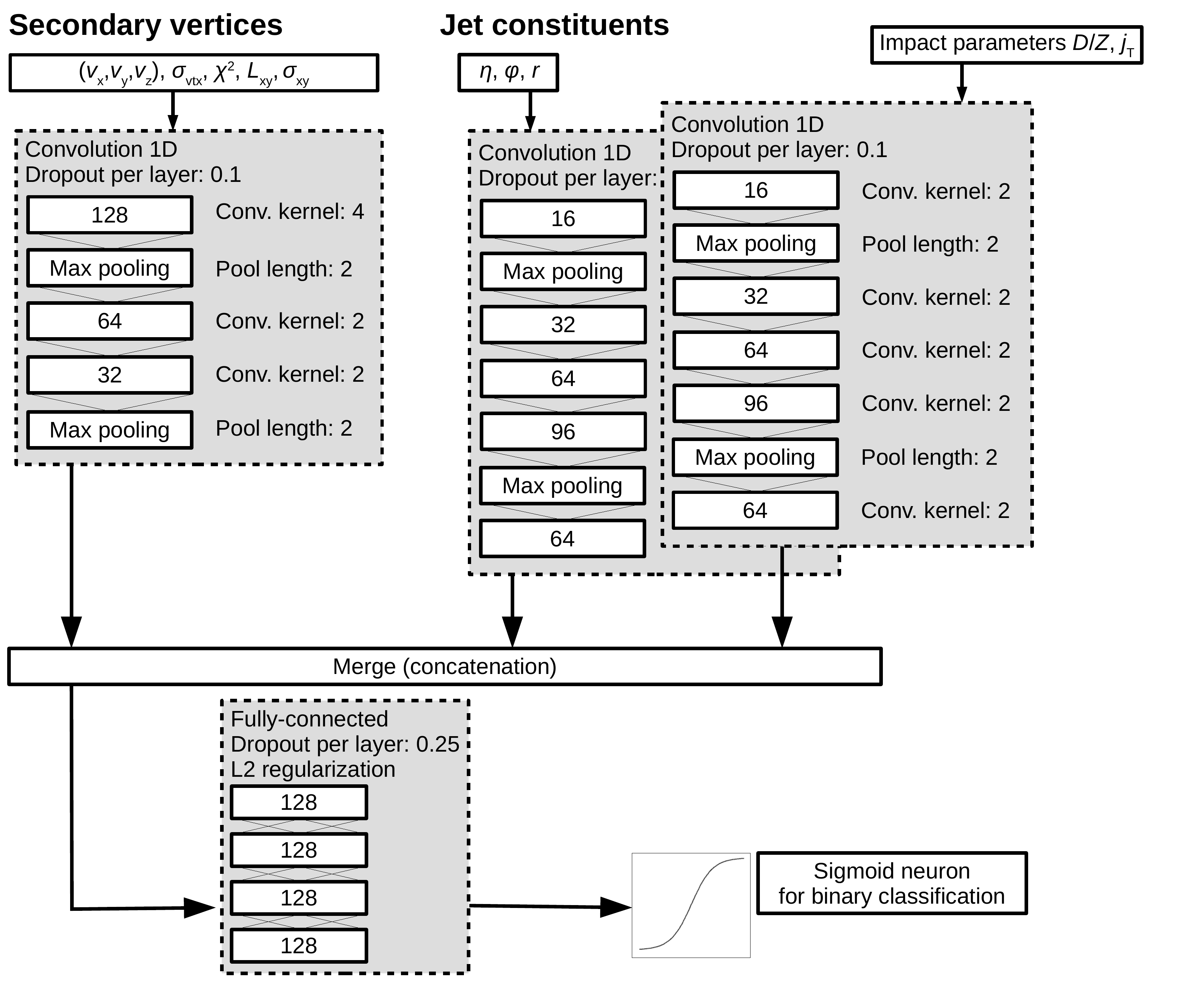}
\centering
\caption{Design of the used model: Three branches -- one processing secondary vertices, and two for the jet constituents -- are merged and further processed by a four-layered fully-connected network with higher dropout. The output is eventually merged to a neuron with sigmoid activation function to be suitable for classification tasks.}
\label{fig:model}
\end{figure}
%%%%%%%%%%%%%%%%%%%%%%%%%%%%%%%%%

The classifier consists of three branches (subnetworks) of multilayered convolutional neural networks (CNNs). On top, a multilayered fully-connected network is processing the merged data. Keras\,\cite{Keras} is used for creation and training of the model.
The model is depicted in Fig.\,\ref{fig:model}. It uses the following input features:
\begin{itemize}
  \item Secondary vertices: relative vertex position: ($v_\mathrm{x}$, $v_\mathrm{y}$, $v_\mathrm{z}$), vertex dispersion $\sigma_\mathrm{vtx}$, vertex fit $\chi^2$, decay length in $xy$-plane $L_\mathrm{xy}$, uncertainty $\sigma_{xy}$.
  \item Jet constituent properties I: coordinates of the tracks, relative to the jet ($\eta$, $\phi$, $r$).
  \item Jet constituent properties II: impact parameters $D$ (transverse), $Z$ (longitudinal), track $j_\mathrm{T}$ ($p_\mathrm{T}$-projection on jet axis).
\end{itemize}
To measure the performance in p--Pb collisions, the Monte Carlo dataset used for the training and evaluation of the performance was created with PYTHIA\,\cite{PYTHIA} and HIJING\,\cite{HIJING} at $\sqrt{s_\mathrm{NN}} = 5.02$ TeV.

The model was trained using 200\,000 jets for each class. During the training, the performance was evaluated on an independent validation dataset of 50\,000 jets per class. The testing dataset used for the analysis consists of roughly 2M udsg-jets, ~500\,000 c-jets, and 580\,000 b-jets. The jets have been reconstructed using FastJet\,\cite{FastJet} with the anti-$k_\mathrm{T}$ algorithm\,\cite{AntiKT} and resolution parameter $R=0.4$.

\subsection{Results}

To evaluate the performance of the present b-tagging algorithm, the mistagging efficiencies have been calculated. The mistagging efficiency represents the fraction of jets of a certain true type that has been wrongly tagged as b-jets. In this context, the b-jet efficiency is the fraction of tagged b-jets out of the true full sample.
The mistagging efficiencies are compared to existing simulation results from ALICE using a cut-based approach\,\cite{CutBasedResults}.

%%%%%%%%%%%%%%%%%%%%%%%%%%%%%%%%%
\begin{figure}[htp]
\includegraphics[width=0.75\textwidth]{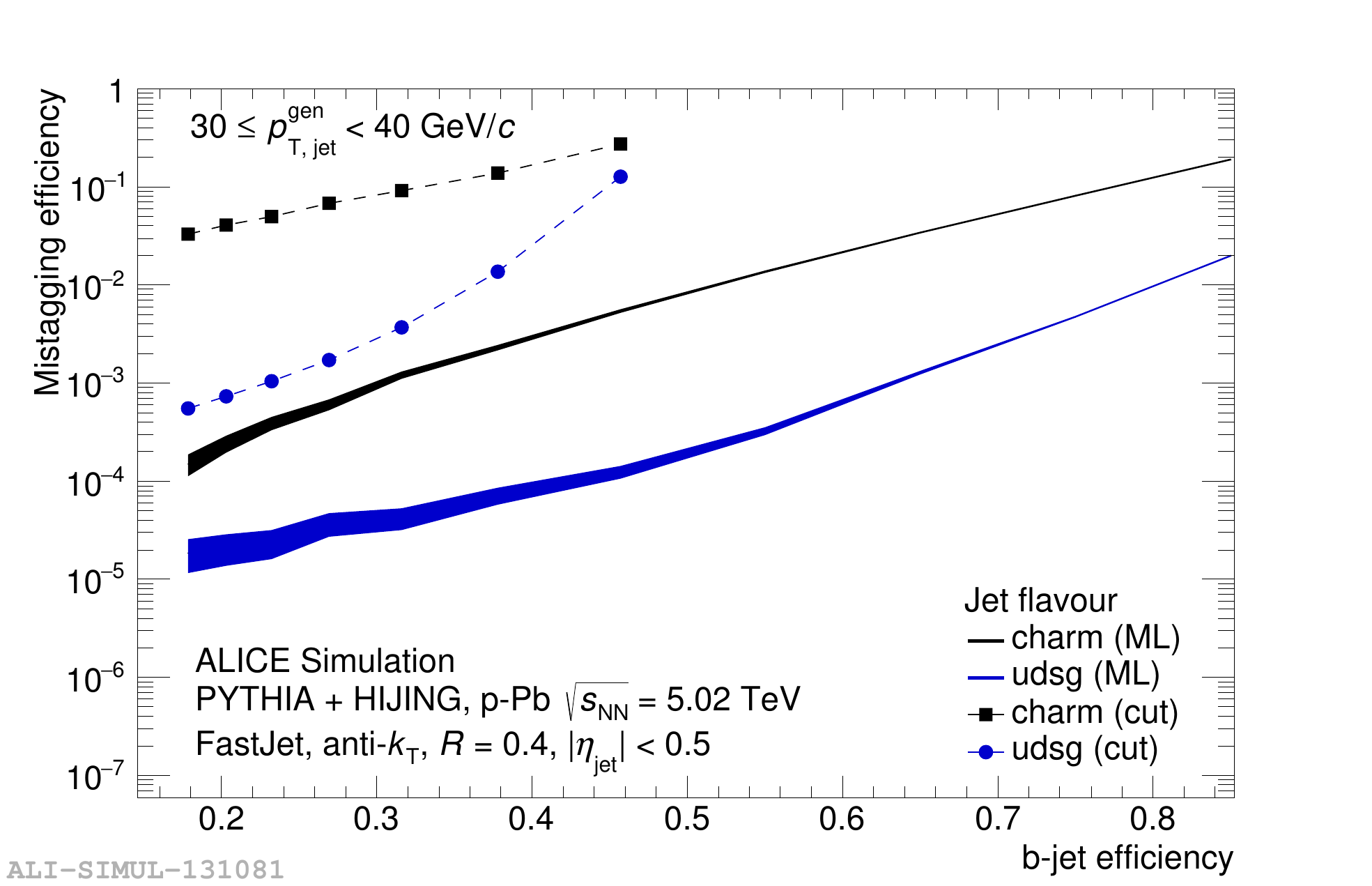}
\centering
\caption{c- and udsg-jets (mis-)tagging efficiencies for several b-jet efficiencies for jets with $30 \leq p_\mathrm{T, jet} < 40$ GeV$/c$ and comparison to cut-based results.}
\label{fig:plot1}
\end{figure}
%%%%%%%%%%%%%%%%%%%%%%%%%%%%%%%%%

%%%%%%%%%%%%%%%%%%%%%%%%%%%%%%%%%
\begin{figure}[htp]
\includegraphics[width=0.75\textwidth]{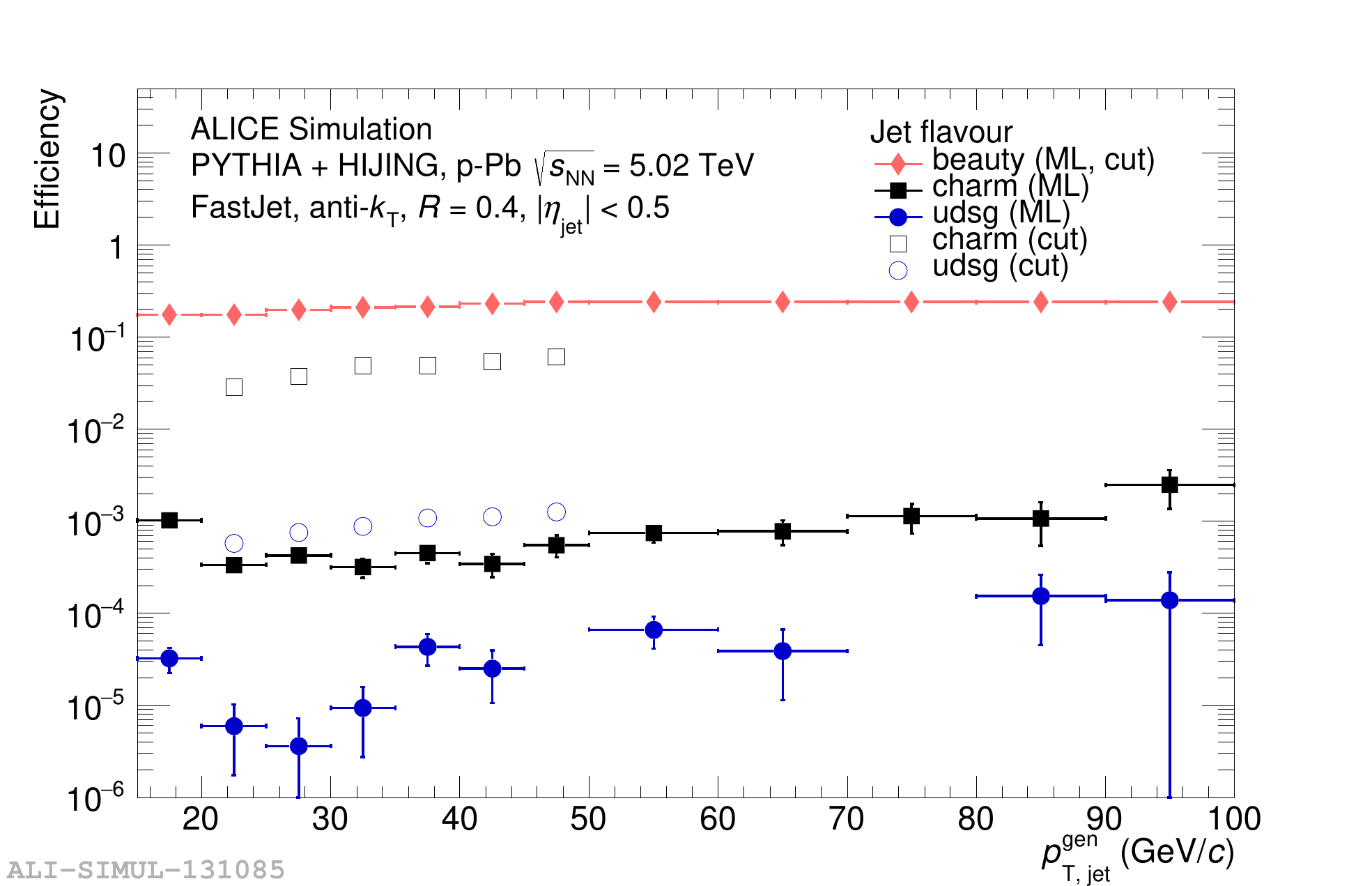}
\centering
\caption{c- and udsg-jets (mis-)tagging efficiencies for given b-jet efficiency (20\%) vs. generated jet $p_\mathrm{T}$ and comparison to cut-based results.}
\label{fig:plot2}
\end{figure}
%%%%%%%%%%%%%%%%%%%%%%%%%%%%%%%%%

\newpage

Figure \ref{fig:plot1} shows the mistagging efficiency for charm and light-flavor jets (udsg) for given b-jet efficiencies and for jets with $30 \leq p_\mathrm{T, jet}^\mathrm{gen} < 40$ GeV/$c$. For both jet types, the mistagging efficiencies are much lower for all b-jet efficiencies than for the standard cut-based method.

%%%%%%%%%%%%%%%%%%%%%%%%%%%%%%%%%
\begin{figure}[htp]
\includegraphics[width=0.75\textwidth]{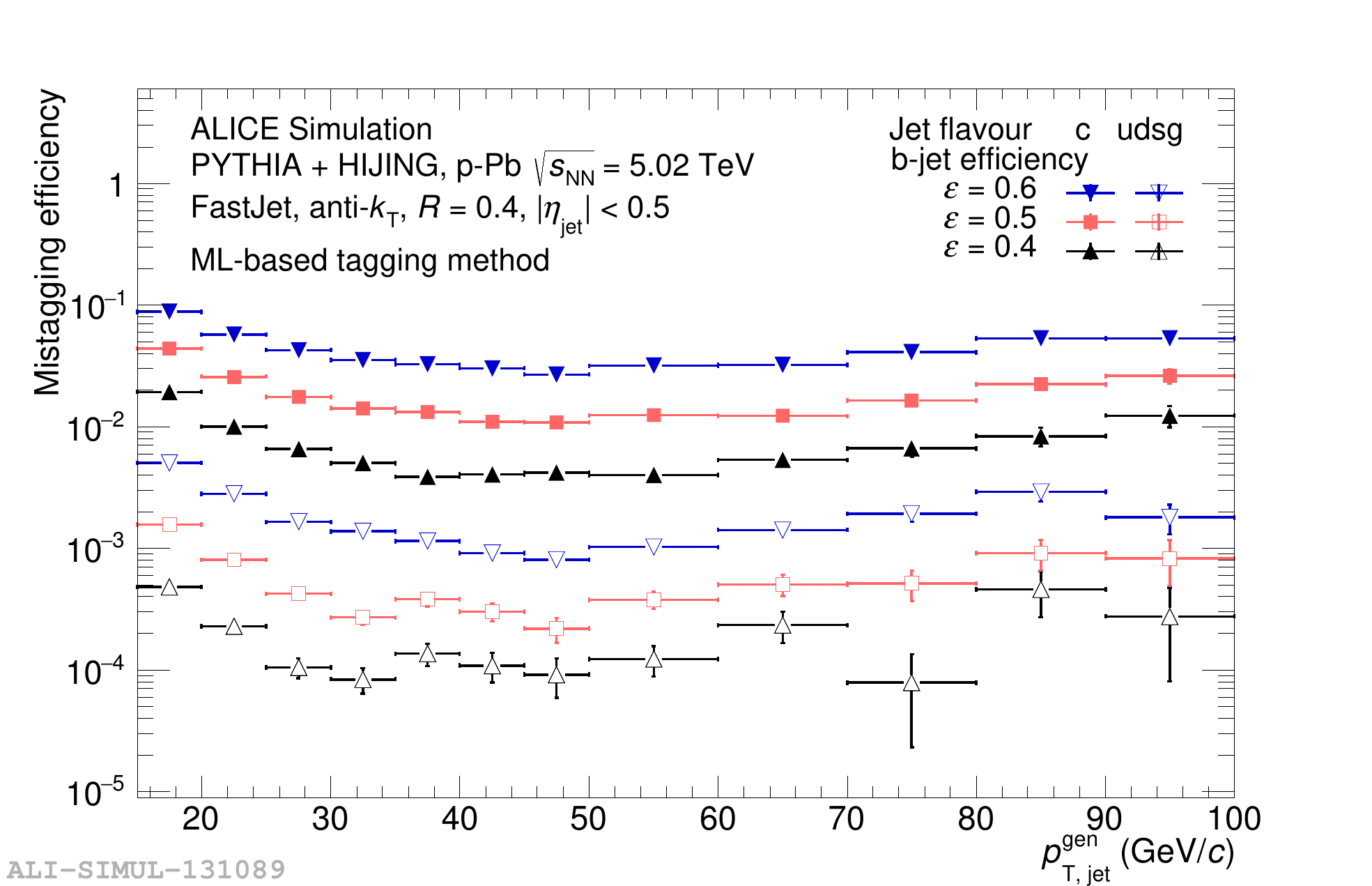}
\centering
\caption{c- and udsg-jets (mis-)tagging efficiencies vs. generated jet $p_\mathrm{T}$ for several efficiencies.}
\label{fig:plot3}
\end{figure}
%%%%%%%%%%%%%%%%%%%%%%%%%%%%%%%%%

This result is consistent with Fig.\,\ref{fig:plot2} that shows the mistagging efficiencies vs. jet transverse momentum for a given b-jet efficiency (roughly 20\%). This figure shows that the present machine-learning-based method has a superior performance compared to the cut-based method.

\newpage

According to the achieved performance in terms of mistagging efficiency, the present method might allow much higher b-jet efficiencies and thus higher statistics in the b-tagged sample. It is important to keep in mind that the mistagging efficiencies, in particular for the light-flavor jets, must be much smaller than the b-jet efficiencies due to the very small fraction of b-jets of 3-5\% in p--Pb collisions\,\cite{bjetfrac}. In Fig.\,\ref{fig:plot3}, the performance for several higher b-jet efficiencies are compared.

\section{Low-mass dielectron identification}

Pairs of electrons and positrons (dielectrons) are created at all stages of a heavy-ion collision. Because they do not interact strongly, their interaction with the medium after creation is negligible. Thus, dielectrons are interesting probes for the Quark-Gluon Plasma. The present analysis focuses on low-mass dielectron pairs\,\cite{LMDPoster}.

\begin{figure}[htp]
\includegraphics[width=0.55\textwidth]{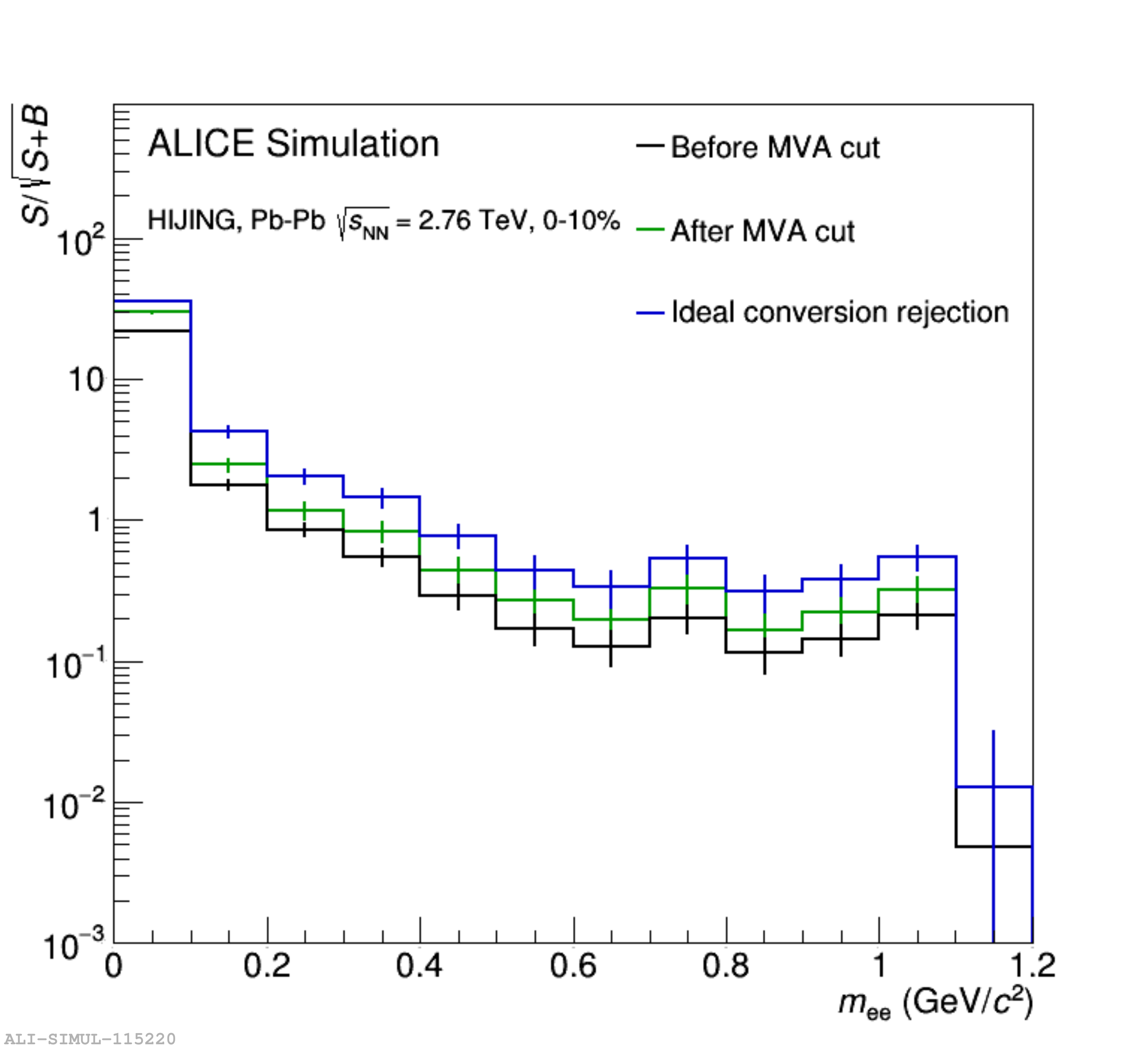}
\centering
\caption{Signal-to-noise ratio before and after the MVA cut with a comparison to the theoretical limit}
\label{fig:plot4}
\end{figure}

A large contribution to the background comes from dielectrons that are formed from one or two photo-conversion electrons. To suppress these two background sources, two fully-connected multilayered networks are trained to select the two types of background electrons. The outputs of both networks are used as cut parameters and the cut is done such that the signal is most significant. In Fig.\,\ref{fig:plot4}, the gain in the signal-to-noise ratio when using the present multivariate analysis (MVA) method -- depicted by the black and green curve -- for HIJING Pb--Pb events is compared to the theoretical limit if all conversion electrons were rejected, which is shown in the blue curve. The result is very promising.

%%%%%%%%%%%%%%%%%%%%%%%%%%%%%%%%%
%%%%%%%%%%%%%%%%%%%%%%%%%%%%%%%%%

In addition to the background rejection method for the dielectrons presented here, the identification of electrons itself can also be improved using MVA techniques such as a Boosted Decision Tree. On this topic, first performance studies have been carried out in pp collisions. Soon, a similar evaluation will be done for Pb--Pb collisions.

\newpage
\section{Summary}
For b-jet tagging, a deep learning tagger has been developed and its performance has been evaluated and compared to the cut-based method. The deep learning tagger allows much higher b-jet efficiencies and lower mistagging rates.
An application to real pp and p--Pb data is ongoing and first results are promising.

For the low-mass dielectron identification, two multilayered neural networks are used to suppress background from photo-conversion electrons. A promising signal-to-noise ratio is expected from Monte Carlo simulations.
The next milestone is to apply this approach to pp, p--Pb, and Pb--Pb data.

\end{document}